\documentclass[unnumsec,webpdf,contemporary,large, numbered,nocrop]{oup-authoring-template}%
\graphicspath{{Fig/}}

\usepackage{graphicx}
\usepackage{amsmath,amssymb,amsfonts}
\usepackage{booktabs}
\usepackage{multirow}
\usepackage{float}
\usepackage{wrapfig}
\usepackage{hyperref}
\usepackage{comment}
\usepackage{algorithm}
\usepackage{algpseudocode}

\usepackage{enumitem}
\setlist{nosep,leftmargin=*}

\theoremstyle{thmstyleone}%
\theoremstyle{thmstyletwo}%
\theoremstyle{thmstylethree}%

\newcommand{\best}[1]{\textbf{#1}}
\makeatletter
\let\ps@opening\ps@plain
\let\ps@headings\ps@plain
\makeatother

\begin{document}

% ==================== Front matter ====================
\journaltitle{ }
\DOI{ }
\copyrightyear{ }
\pubyear{ }
%\vol{ }
%\issue{ }
\access{ }
\appnotes{ }

\firstpage{1}

%\title[Ultrametric-aware mtDNA phylogeny reconstruction]{Phylogeny reconstruction\\
%from partially computed mtDNA distance matrices\\
%via ultrametric-aware completion}

\title[Ultrametric-aware mtDNA phylogeny reconstruction]{Hyb-Adam-UM: hybrid ultrametric-aware mtDNA phylogeny reconstruction}

\author[1]{Dmitrii Chaikovskii}
\author[2]{Weilai Qu}
\author[2]{Boris Melnikov}
\author[1,3]{Ye Zhang}
\author[4,$\ast$]{Yuehong Zhao}

\address[1]{\orgdiv{MSU-BIT-SMBU Joint Research Center of Applied Mathematics},
\orgname{Shenzhen MSU-BIT University},
\orgaddress{\postcode{518172}, \state{Shenzhen}, \country{China}}}

\address[2]{\orgdiv{Faculty of Computational Mathematics and Cybernetics},
\orgname{Shenzhen MSU-BIT University},
\orgaddress{\postcode{518172}, \state{Shenzhen}, \country{China}}}

\address[3]{\orgdiv{School of Mathematics and Statistics},
\orgname{Beijing Institute of Technology},
\orgaddress{\postcode{100081}, \state{Beijing}, \country{China}}}

\address[4]{\orgdiv{School of Medicine},
\orgname{Chinese University of Hong Kong},
\orgaddress{\postcode{518172}, \state{Shenzhen}, \country{China}}}

\corresp[$\ast$]{Corresponding author, E-mail: \href{mailto:zhaoyuehong94@foxmail.com}{zhaoyuehong94@foxmail.com}}

%\received{Date}{0}{Year}
%\revised{Date}{0}{Year}
%\accepted{Date}{0}{Year}

\abstract{
\textbf{Motivation:} mtDNA distance matrices are standard inputs for distance-based phylogeny, but computing all pairwise alignments is costly; missing entries can degrade topology and branch lengths, and generic completion may break tree-like (ultrametric) structure.\\
\textbf{Results:} We propose Hyb-Adam-UM, which starts from an alignment-limited Needleman--Wunsch distance backbone and completes the matrix by minimizing a robust triplet ultrametric-violation functional. An Adam-style finite-difference optimizer updates only missing entries and enforces symmetry, non-negativity, and a zero diagonal. From one complete $15\times 15$ reference matrix we generate 20 masked instances at 30\%, 50\%, 65\%, and 85\% missingness. Hyb-Adam-UM consistently reduces ultrametric violations and delivers competitive reconstruction error together with improved topological accuracy and branch-length agreement relative to MW$^\star$-/NJ$^\star$-projection (which exactly preserve observed distances) and Soft-Impute, with the most pronounced gains at 85\% missingness.\\
\textbf{Availability and implementation:} https://github.com/mitichya/hyb-adam-um/; Zenodo: https://doi.org/10.5281/zenodo.18609748.\\
\textbf{Supplementary information:} Supplementary data available online.
}


\maketitle

\thispagestyle{plain}
\pagestyle{plain}

% ==================== Main text ====================
\section{Introduction}

Mitochondrial DNA (mtDNA) is a central data source in evolutionary biology, population genetics, and phylogenetics due to maternal inheritance, limited recombination, and relatively high mutation rates. These properties make mtDNA particularly useful for tracing maternal lineages, estimating divergence times, and reconstructing phylogenetic relationships among closely related species \cite{BallardWhitlock2004}. In many workflows, a pairwise distance matrix derived from mtDNA sequences is not only a summary representation of sequence divergence, but also the \emph{direct input} to distance-based phylogenetic \emph{tree reconstruction} procedures such as Neighbor-Joining and UPGMA \cite{SaitouNei1987,SokalMichener1958}. Therefore, the quality of the distance matrix has immediate consequences for inferred tree topology and branch lengths.

The precision and reliability of distance-based inference depend strongly on the \emph{completeness} and \emph{global consistency} of the distance matrix. At the same time, building a complete matrix is computationally demanding: for $n$ taxa, one must compute $n(n-1)/2$ pairwise distances. When distances are computed via sequence alignment, each entry may require dynamic programming over sequence lengths, as in classical global alignment \cite{NeedlemanWunsch1970}. In our practical setting (human-length mtDNA of approximately $16{,}569$ base pairs \cite{Anderson1981}), a Needleman--Wunsch global alignment can take on the order of minutes per pair in a straightforward implementation, which scales into hours or days for matrices of moderate size and into months for larger taxon sets. For example, a $30\times 30$ matrix has $435$ off-diagonal distances; if each distance required only three minutes, the full matrix would require roughly one day of wall-clock time on a modern computer. If one considers hundreds of taxa, complete matrix computation becomes infeasible without extensive parallel computing. Since distance-based methods typically assume a fully specified matrix (or behave unpredictably under missing entries), incomplete computation can translate into unstable or biased reconstructed trees.

As a consequence, distance matrices used in practice are often incomplete. Incompleteness may arise because only a subset of pairwise alignments is computed (compute-budget constraints), because some sequences are missing or rejected by quality filters, or because certain alignments fail. Unfortunately, missing entries can severely affect downstream phylogenetic inference: even when tree construction remains feasible, naive handling (e.g.\ pairwise deletion or simplistic imputation) may introduce bias, distort branch lengths, and reduce resolution \cite{WiensMorrill2011}. Therefore, robust restoration of incomplete distance matrices is an important enabling step for large-scale comparative genomics and \emph{reliable reconstruction of phylogenetic trees}.

A further complication is that mtDNA distance matrices are not arbitrary numerical arrays. Although the entries are symmetric and nonnegative, the triangle inequality may be violated. Nevertheless, they often preserve a strong tree-like, approximately ultrametric signal: in many triplets, two distances are nearly equal and exceed the third.
 Generic matrix completion methods (e.g.\ low-rank completion) can fill missing values effectively in a least-squares sense \cite{CandesRecht2009,MazumderHastieTibshirani2010}, yet may fail to preserve hierarchical constraints that are important for phylogenetic interpretability.

\subsection{Contributions}
%This paper targets \emph{phylogenetic tree reconstruction from partially computed distance matrices} by designing a completion procedure that preserves tree-like geometry rather than only minimizing entrywise error. Concretely, we introduce a hybrid method that couples biological signals from alignment with an explicit global-consistency objective:
This paper addresses \emph{mtDNA-based phylogenetic tree reconstruction from mitochondrial DNA sequences} in settings where only a subset of pairwise distances can be computed. We propose a distance-matrix completion procedure that promotes tree-consistent geometry rather than optimizing only entrywise reconstruction error. Concretely, we introduce a hybrid method that couples biological signals from alignment with an explicit global-consistency objective:
\begin{enumerate}
\item \textbf{Warm-start via alignment:} we use the available Needleman–Wunsch distances \cite{NeedlemanWunsch1970} under a limited alignment budget as an observed backbone and initialize the remaining missing entries (e.g., by the mean of observed distances).
\item \textbf{Ultrametric-aware optimization:} we complete the remaining missing entries by minimizing a triplet-based ultrametric-violation functional, using an Adam-style optimizer \cite{KingmaBa2014} with finite-difference gradients, while enforcing symmetry, non-negativity, and a zero diagonal.
\end{enumerate}
Beyond matrix-level accuracy, we evaluate downstream Neighbor-Joining tree agreement with the reference (normalized Robinson--Foulds distance and patristic-distance agreement), showing that ultrametric-aware completion translates into improved tree topology and branch-length consistency under high missingness.
Starting from a single complete $15\times 15$ mtDNA reference distance matrix, we generate $20$ incomplete instances via symmetric random masking at missingness levels of 30\%, 50\%, 65\%, and 85\% (five independent masks per level), and benchmark against three baselines: MW$^\star$-proj \cite{MakarenkovLapointe2004}, NJ$^\star$-proj (Neighbor-Joining-based completion) \cite{SaitouNei1987}, and low-rank matrix completion via Soft-Impute \cite{MazumderHastieTibshirani2010}.

\section{Preliminaries}

%\subsection{Distance matrices and triangle constraints}

\begin{wrapfigure}[9]{cr}{0.6\linewidth}
\vspace{-7.5mm}
\centering
\includegraphics[width=0.25\textwidth]{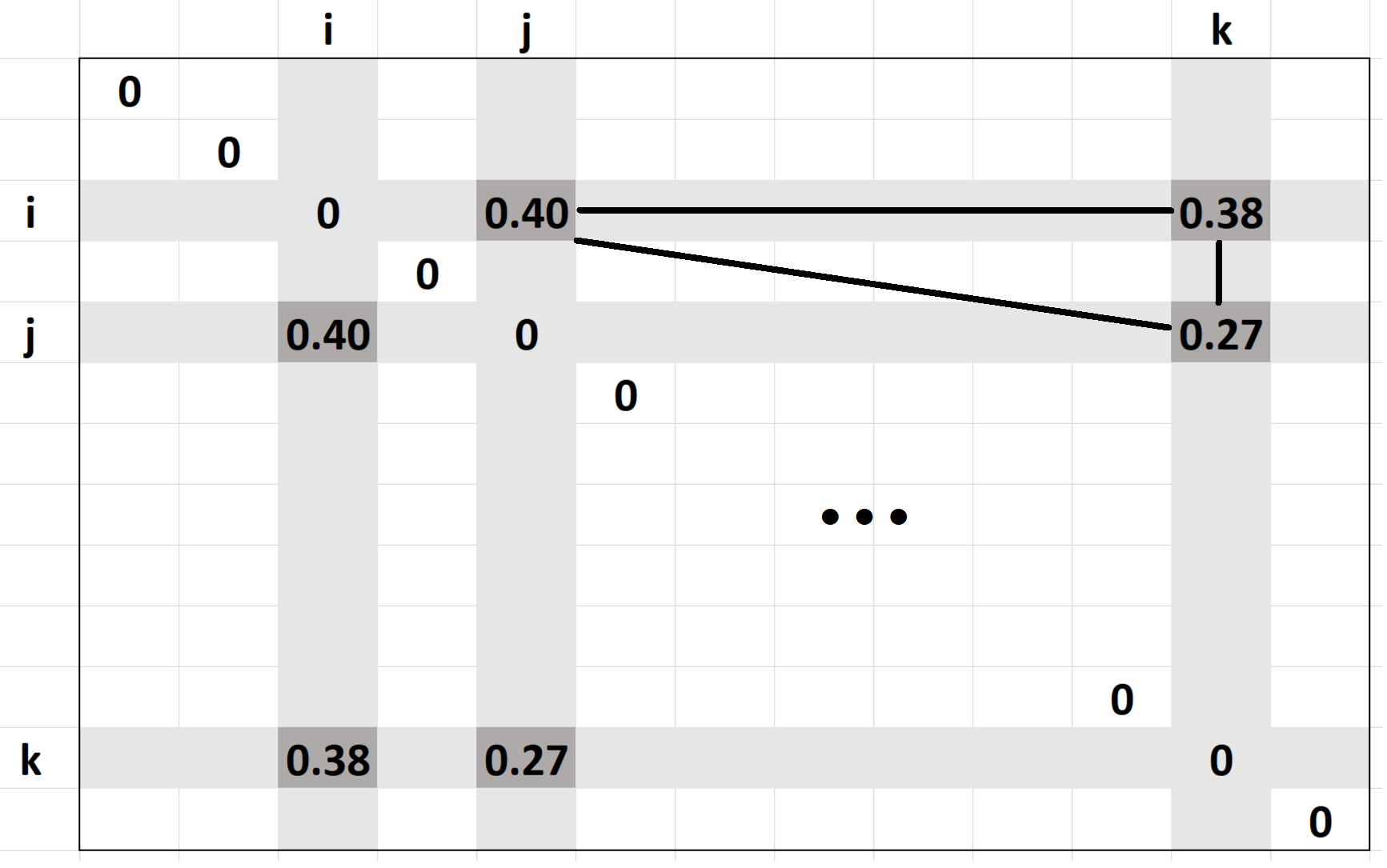}
\vspace{-5mm}
\caption{Illustration of triangle relations arising from a symmetric distance matrix.}
\label{fig:matrica}

\end{wrapfigure}

Let $D\in\mathbb{R}^{n\times n}$ be a symmetric distance matrix with
\[
D_{ii}=0,\qquad D_{ij}=D_{ji}\ge 0,\qquad 1\le i,j\le n.
\]
For any triple of indices $(i,j,k)$, the three distances $(D_{ij},D_{ik},D_{jk})$ ideally form a non-degenerate triangle satisfying the triangle inequality (Figure \ref{fig:matrica}). When distances approximate divergence times (a molecular-clock regime), many triplets exhibit a “short-base” isosceles pattern: two closely related taxa have a small mutual distance, while both are at similar, larger distance from a third, more distantly related taxon.

A classical motivating example involves three taxa: human (H), chimpanzee (C), and bonobo (B). Biological consensus suggests that (i) the ancestral line of humans diverged from the common ancestor of chimpanzees/bonobos earlier, and (ii) chimpanzees and bonobos diverged more recently. In a distance matrix, this implies that two distances (H--C and H--B) are close and larger than the remaining distance (C--B), yielding an isosceles triangle with a short base. Precise divergence times are not required for our analysis; instead, we exploit the characteristic triangle geometry induced by near-ultrametric distances and use it to define an ultrametric consistency objective.

%\subsection{Triplet violation measures and a global violation score}

\begin{wrapfigure}[9]{cr}{0.4\linewidth}
\vspace{-6mm}
\centering
\includegraphics[width=0.11\textwidth]{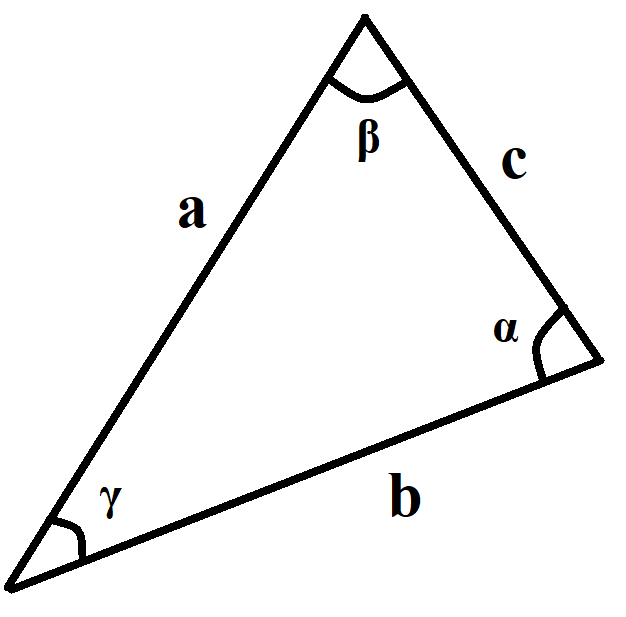}
\caption{Triangle with side lengths $a,b,c$ and angles $\alpha,\beta,\gamma$.}
\label{fig:triangle_abc}
\vspace{-2mm}
\end{wrapfigure}

For a given triangle with side lengths $a\ge b\ge c$, numerous violation measures can be defined to quantify deviation from an ultrametric-like triangle (Figure \ref{fig:triangle_abc}). In our previous mtDNA-related works \cite{DNA-0,DNA-1,DNA-2,DNA-3,DNA-4,DNA-5,DNA-6}, we explored several variants. Table~\ref{tab:badness} lists representative definitions and example values.

\begin{table}[H]
\caption{Examples of triplet violation measures for triangles and pseudo-triangles. Angles are in degrees (rounded), with $a\ge b\ge c$ and $\alpha\ge\beta\ge\gamma$.}
\label{tab:badness}
\centering

%\setlength{\tabcolsep}{3pt}      % tighter columns (default ~6pt)
%\renewcommand{\arraystretch}{1.0} % slightly tighter rows

%\resizebox{0.48\textwidth}{!}{%
\begin{tabular}{cc||ccccc}
\toprule
\textbf{Sides$^{1,2}$} & \textbf{Angles} &
\textbf{A.(0)} &
\textbf{A.(1)} &
\textbf{A.(2)} &
\textbf{S.(3)} &
\textbf{S.(4)} \\
$a,b,c$ & $\alpha,\beta,\gamma$ &
$\frac{\alpha-\beta}{\gamma}$ &
$\frac{\alpha-\beta}{180} $ &
$\frac{\alpha-\beta}{\alpha} $ &
$\frac{a-b}{a} $ &
$\frac{a-b}{c}$ \\
\midrule
$1,1,1$ & $60,60,60$ & $0$ & $0$ & $0$ & $0$ & $0$ \\
$5,5,4$ & $66,66,47$ & $0$ & $0$ & $0$ & $0$ & $0$ \\
$42,41,28$ & $72,68,39$ & $0.10$ & $0.04$ & $0.05$ & $0.02$ & $0.04$ \\
$19,18,17$ & $66,60,55$ & $0.11$ & $0.07$ & $0.09$ & $0.05$ & $0.06$ \\
$10,9,8$ & $72,59,50$ & $0.26$ & $0.14$ & $0.18$ & $0.10$ & $0.13$ \\
$6,5,5$ & $74,53,53$ & $0.39$ & $0.23$ & $0.28$ & $0.17$ & $0.20$ \\
$13,12,5$ & $90,67,23$ & $1.00$ & $0.25$ & $0.25$ & $0.08$ & $0.20$ \\
$5,4,3$ & $90,53,37$ & $1.00$ & $0.41$ & $0.41$ & $0.20$ & $0.33$ \\
\midrule
$12,6,5$ & $-$ &  &  & $1.09$ &  &  \\
$20,6,5$ & $-$ &  &  & $1.81$ &  &  \\
\bottomrule
\end{tabular}%

\vspace{2pt} % tight space between table and notes

{\footnotesize
\setlength{\parskip}{0pt}
\setlength{\parindent}{0pt}
\textsuperscript{1} Angles are rounded to the nearest degree, so the sum may differ from $180^\circ$.\\[-1pt]
\textsuperscript{2} $a\geqslant b\geqslant c$, $\alpha\geqslant\beta\geqslant\gamma$.\\[-1pt]
}

\end{table}

In this paper, we use an angle-based measure closely aligned with the angle-gap ratio A.(0) (with additional robustness for triangle-inequality violations) and sum it over all triplets to obtain a global violation score. The total number of triplets is $\binom{n}{3}$. Even for moderate $n$, the number of triplets is large, e.g.\ $\binom{34}{3}=5984$, and for $n=15$ (our experimental setting) $\binom{15}{3}=455$.

\subsection{Robust triplet penalty and $\Delta$ functional}

For each triplet $(i,j,k)$, define a robust penalty $\delta(i,j,k)$ computed from the three side lengths. If triangle inequality is severely violated, we assign a large penalty; otherwise we compute angles by the law of cosines and quantify separation between the two largest angles normalized by the smallest. The overall score is
\[
\Delta(D)=\sum_{i<j<k}\delta(i,j,k).
\]
Because $\Delta$ scales with the number of triplets, we also report the per-triplet score
\begin{equation}\label{eq:delta_normalized}
\overline{\Delta}(D)=\frac{\Delta(D)}{\binom{n}{3}},
\end{equation}
which is directly comparable across different $n$. The robust computation of $\delta(i,j,k)$ and the accumulation of $\Delta(D)$ are summarized in Algorithm~\ref{alg:delta_triplet_robust}.

\begin{algorithm}[H]
\caption{Robust computation of $\delta(i,j,k)$ and total $\Delta(D)$}
\label{alg:delta_triplet_robust}
\begin{algorithmic}[1]
\Require Distance matrix $D$, constant $\omega=2$, small $\varepsilon>0$ (e.g.\ $10^{-8}$)
\Ensure Total score $\Delta(D)$
\Function{Clamp}{$x$}
  \State \Return $\min(1,\max(-1,x))$
\EndFunction
\State $\Delta \gets 0$
\For{each triplet $(i,j,k)$ with $i<j<k$}
  \State $a\gets D_{ij}$,\; $b\gets D_{ik}$,\; $c\gets D_{jk}$; sort so that $a\ge b\ge c$
  \If{$a \ge b + c$} \Comment{Severe triangle-inequality violation}
    \State $\delta \gets \max\!\left(\dfrac{a}{\max(b+c,\varepsilon)},\; \omega\right)$
  \Else
    \State $\cos\alpha \gets \Call{Clamp}{\dfrac{b^2+c^2-a^2}{2bc+\varepsilon}}$
    \State $\cos\beta  \gets \Call{Clamp}{\dfrac{a^2+c^2-b^2}{2ac+\varepsilon}}$
    \State $\cos\gamma \gets \Call{Clamp}{\dfrac{a^2+b^2-c^2}{2ab+\varepsilon}}$
    \State $\alpha\gets\arccos(\cos\alpha)$;\; $\beta\gets\arccos(\cos\beta)$;\; $\gamma\gets\arccos(\cos\gamma)$
    \State Sort angles so that $A\ge B\ge \Gamma$ where $\{A,B,\Gamma\}=\{\alpha,\beta,\gamma\}$
    \State $\delta \gets \dfrac{A-B}{\max(\Gamma,\varepsilon)}$
  \EndIf
  \State $\Delta \gets \Delta + \delta$
\EndFor
\State \Return $\Delta$
\end{algorithmic}
\end{algorithm}

\section{Method}

\subsection{Problem formulation}

\begin{comment}
Let $D_{\mathrm{ref}}\in\mathbb{R}^{n\times n}$ denote a complete reference distance matrix derived from mtDNA alignments. In realistic settings, only a subset of entries is observed, producing an incomplete matrix $D_{\mathrm{obs}}$. The task is to estimate a completed matrix $\widehat D$ that (i) preserves the observed distances, (ii) satisfies the distance-matrix constraints (symmetry, non-negativity, and zero diagonal), and (iii) is globally consistent with hierarchical divergence patterns, as quantified by a small value of $\Delta(\widehat D)$ (or $\overline{\Delta}(\widehat D)$).

We adopt an explicit optimization formulation:
\begin{equation}\label{eq:opt}
\min_{D\in\mathbb{R}^{n\times n}} \ \Delta(D)
\quad \text{s.t.}\quad
\begin{cases}
D_{ii}=0,\quad D_{ij}=D_{ji}\ge 0,\\
D_{ij}=D_{\mathrm{obs},ij},\ \ (i,j)\in\Omega,
\end{cases}
\end{equation}
where $\Omega$ denotes the set of observed entries.

\end{comment}

Let $\{s_i\}_{i=1}^n$ denote the mtDNA sequences for $n$ taxa. Pairwise distances are obtained from Needleman–Wunsch (NW) global alignments; however, in practice only a subset of alignments is computed due to compute-budget constraints. This yields an \emph{incomplete, partially computed} distance matrix $D_{\mathrm{obs}}\in\mathbb{R}^{n\times n}$ with
\[
(D_{\mathrm{obs}})_{ii}=0,\qquad (D_{\mathrm{obs}})_{ij}=(D_{\mathrm{obs}})_{ji}\ge 0\ \text{for}\ (i,j)\in\Omega,
\]
and missing values for $(i,j)\notin\Omega$, where $\Omega\subset\{(i,j):1\le i<j\le n\}$ is the set of pairs whose distances were actually computed by NW. The completion task is to estimate a full matrix $\widehat D$ that (i) preserves all alignment-derived observed distances, (ii) satisfies distance-matrix constraints (symmetry, non-negativity, and zero diagonal), and (iii) is globally consistent with hierarchical divergence patterns, as quantified by a small value of $\Delta(\widehat D)$ (or $\overline{\Delta}(\widehat D)$).

We formulate ultrametric-aware completion as a constrained optimization problem
\begin{equation}\label{eq:opt}
\min_{D\in\mathbb{R}^{n\times n}} \ \Delta(D)
\quad \text{s.t.}\quad
\begin{cases}
D_{ii}=0,\quad D_{ij}=D_{ji}\ge 0,\\
D_{ij}=(D_{\mathrm{obs}})_{ij},\ \ (i,j)\in\Omega.
\end{cases}
\end{equation}
For evaluation only (not as an input), we denote by $D_{\mathrm{ref}}$ a complete reference matrix computed by performing NW alignments for all $\binom{n}{2}$ pairs; it serves as the ground truth for reporting reconstruction error in the experiments.

\subsection{Hybrid strategy}

Our method proceeds in two stages: we initialize the unobserved entries of a partially observed alignment-derived distance matrix, and then refine those entries by minimizing a global triplet-based ultrametric-violation score while keeping observed distances fixed.

\paragraph{Stage 1: observed-alignment backbone and initialization.}

We start from a partially computed mtDNA distance matrix $D_{\mathrm{obs}}$, where the available entries are obtained from Needleman--Wunsch (NW) global alignments and the remaining entries are unobserved due to a limited alignment budget (or alignment failures). In our experiments, $D_{\mathrm{obs}}$ is generated by masking a fully computed reference matrix $D_{\mathrm{ref}}$ to simulate this partial-observation setting.

\paragraph{Stage 2: ultrametric-aware completion by Adam optimization.}
We treat the remaining missing entries as free variables and minimize $\Delta(D)$ iteratively. Since $\Delta(D)$ is non-smooth, we estimate gradients by symmetric finite differences and use an Adam-style optimizer, updating \emph{only} the missing entries while keeping observed distances fixed.

Let $M:=\{(i,j):\, i<j,\ (i,j)\ \text{is missing}\}$ denote the set of missing index pairs in the upper triangle; by symmetry we enforce $D_{ji}=D_{ij}$ for all $(i,j)\in M$.
At iteration $t$, we maintain a current matrix $D_t$. For each $(i,j)\in M$, we approximate the gradient by
\begin{equation}\label{eq:gradient}
[g_t]_{ij}\approx \frac{\Delta\!\left(D_t+h(e_{ij}+e_{ji})\right)-\Delta\!\left(D_t-h(e_{ij}+e_{ji})\right)}{2h},
\  [g_t]_{ji}=[g_t]_{ij},
\end{equation}
where $e_{ij}$ is the $(i,j)$-th standard basis matrix and $h$ is small (we use $h=5\times 10^{-5}$). Observed entries are fixed throughout the optimization.

Optionally, we apply weight decay on missing entries only,
\begin{equation}\label{eq:wd}
[g_t]_{ij} \leftarrow [g_t]_{ij} + \lambda D_{t,ij}, \qquad (i,j)\in M,
\end{equation}
although in the experiments reported here we set $\lambda=0$.

We employ Adam moment estimates with standard bias correction,
\begin{align}\label{eq:adamparams}
m_t &= \beta_1 m_{t-1} + (1-\beta_1)\,g_t, &
v_t &= \beta_2 v_{t-1} + (1-\beta_2)\,g_t^{2}, \\
\hat m_t &= \frac{m_t}{1-\beta_1^t}, &
\hat v_t &= \frac{v_t}{1-\beta_2^t},
\end{align}
where $g_t^2$ is taken elementwise (restricted to $(i,j)\in M$). Updates are applied \emph{only} to missing entries:
\begin{equation}\label{eq:update}
D_{t+1}[M] = D_t[M] - \frac{\alpha_t}{\sqrt{\hat v_t[M]}+\varepsilon}\,\hat m_t[M],
\end{equation}
with $(\beta_1,\beta_2,\varepsilon)=(0.9,0.999,10^{-8})$. After each update we enforce symmetry, non-negativity, and a zero diagonal (projection onto the distance-matrix constraint set). In our implementation we additionally apply gradient clipping on missing entries and use a decreasing learning-rate $\alpha_t$ schedule; details are given in the Experiments section.
The overall optimization procedure is summarized in Algorithm~\ref{alg:hyb_adam_um_short_refeq}.

\begin{algorithm}[H]
\caption{Hyb-Adam-UM: ultrametric-aware completion by finite-difference Adam}
\label{alg:hyb_adam_um_short_refeq}
\begin{algorithmic}[1]
\Require Incomplete symmetric matrix $D_{\mathrm{obs}}$ (missing entries flagged), epochs $T$, step size schedule $\{\alpha_t\}$, finite-difference step $h$, Adam $(\beta_1,\beta_2,\varepsilon)$, optional weight decay $\lambda$, optional gradient-clip threshold $c$
\Ensure Completed matrix $\widehat D$ and final score $\Delta(\widehat D)$

\State Define observed set $\Omega:=\{(i,j): (D_{\mathrm{obs}})_{ij}\ \text{observed},\ i<j\}$ and missing set $M:=\{(i,j): i<j,\ (i,j)\notin\Omega\}$
\State Initialize $D\gets D_{\mathrm{obs}}$ and set $D_{ij}$ for $(i,j)\in M$ to the mean of observed off-diagonal distances
\State Initialize Adam moments $m\gets 0$, $v\gets 0$ on indices in $M$
\For{$t=1,\dots,T$}
  \State Compute $\Delta(D)$ using Algorithm~\ref{alg:delta_triplet_robust}
  \State Compute finite-difference gradient $g$ on missing entries via Eq.~\eqref{eq:gradient}
  \If{$\lambda>0$}
    \State Apply weight decay on missing entries via Eq.~\eqref{eq:wd}
  \EndIf
  \If{$c>0$}
    \State Clip the gradient on $M$ (e.g., enforce $\|g\|_2\le c$)
  \EndIf
  \State Update Adam moments and bias correction via Eq.~\eqref{eq:adamparams}
  \State Update missing entries via Eq.~\eqref{eq:update} using $\alpha_t$
  \State Enforce symmetry, keep $D[\Omega]$ fixed
\EndFor
\State \Return $\widehat D\gets D$, $\Delta(\widehat D)$
\end{algorithmic}
\end{algorithm}

\section{Experiments}

%We start from $n=15$ monkey mtDNA sequences retrieved from GenBank (NCBI)\cite{sayers2023ncbi,ncbi_genbank_overview}. The choice $n=15$ yields meaningful triplet statistics ($\binom{15}{3}=455$) while keeping the completion problem non-trivial under substantial missingness.
%Pairwise distances are defined from Needleman--Wunsch (NW) global alignments \cite{NeedlemanWunsch1970} using a fixed scoring scheme and converted into a symmetric non-negative distance (e.g., normalized mismatch/gap rate). 

We consider $n=15$ monkey mitochondrial DNA (mtDNA) sequences retrieved from GenBank (NCBI) \cite{sayers2023ncbi,ncbi_genbank_overview}. To avoid ambiguous alignments and unreliable distance estimates, we exclude sequences that contain unresolved bases (\texttt{N}) before computing pairwise distances. The choice $n=15$ yields sufficiently rich triplet statistics for ultrametricity-based objectives ($\binom{15}{3}=455$) while keeping matrix completion non-trivial under substantial missingness.

Pairwise dissimilarities are computed from Needleman--Wunsch (NW) global alignments (see \cite{NeedlemanWunsch1970}) using a fixed linear scoring scheme: match $+5$, mismatch $-4$, and a symmetric linear gap penalty $-4$ per inserted/deleted position (no affine gap open/extend terms and no substitution matrix). For each pair $(i,j)$, NW produces an optimal aligned pair $(\hat x,\hat y)$ of length $L$, and we define a symmetric non-negative distance by the normalized column-wise disagreement rate,
\[
d_{ij} \;=\; \frac{1}{L}\sum_{k=1}^{L}\mathbf{1}\{\hat x_k \neq \hat y_k\},
\]
where $\mathbf{1}\{\cdot\}$ is the indicator function (equal to $1$ if the condition holds and $0$ otherwise). Gaps (\texttt{-}) are treated as ordinary symbols and therefore contribute to disagreement. The resulting matrix $D_{\mathrm{ref}}\in\mathbb{R}^{15\times 15}$ is symmetric with $d_{ii}=0$, and serves as the reference complete distance matrix prior to applying controlled missingness masks in subsequent experiments.

To model realistic compute-budget constraints, we simulate \emph{partially computed} distance matrices by restricting the set of taxon pairs for which an NW alignment is available.
Let $\Omega_{\triangle}:=\{(i,j):1\le i<j\le n\},\ |\Omega_{\triangle}|=\binom{n}{2},$ denote the set of unique off-diagonal taxon pairs (one representative per symmetric pair).
For each missingness level $p\in\{30\%,50\%,65\% ,85\%\}$ and replicate $r=1,\dots,5$, we sample an observed pair set $\Omega_{p,r}\subset\Omega_{\triangle}$ with $|\Omega_{p,r}|\approx (1-p)\binom{n}{2}$.
The incomplete matrix $D_{\mathrm{obs}}^{(p,r)}$ is defined for $i<j$ as
\[
(D_{\mathrm{obs}}^{(p,r)})_{ij}=
\begin{cases}
D_{\mathrm{ref},ij}, & (i,j)\in\Omega_{p,r},\\
\text{missing}, & (i,j)\notin\Omega_{p,r},
\end{cases}
\]
and then extended by $D_{ji}=D_{ij}$ and $D_{ii}=0$.

Here $D_{\mathrm{ref}}$ is used only to provide the numerical values of the NW distances for the sampled observed pairs and to enable objective evaluation; conceptually, $(i,j)\in\Omega_{p,r}$ corresponds to pairs whose NW alignments were actually computed under a limited budget.
Thus, the observed NW distances form a biologically informed \emph{warm-start backbone}, and completion is required only for uncomputed pairs.

%To model realistic compute-budget constraints, we simulate \emph{partially computed} distance matrices by limiting the number of NW alignments.For each missingness level $p\in\{30\%,50\%,65\%\}$, from $D_{\mathrm{ref}}$, we generate $5$ incomplete matrices by randomly masking off-diagonal entries while preserving symmetry, i.e., whenever $(i,j)$ is masked, $(j,i)$ is masked as well, and the diagonal remains zero.  Thus we total obtain $5+5+5=15$ incomplete instances. Thus, the NW-computed entries provide a biologically informed \emph{warm-start} (the observed backbone), and completion is required only for truly uncomputed pairs.

\subsection{Completion methods and hyperparameters}

We apply Hyb-Adam-UM (Algorithm~\ref{alg:hyb_adam_um_short_refeq}) to each incomplete matrix.
We used $T=3000$ epochs, central-difference gradients with $h=5\times10^{-5}$, and Adam parameters $(\beta_1,\beta_2,\varepsilon)=(0.9,0.999,10^{-8})$.
Missing entries were initialized to the mean of observed off-diagonal distances and projected to non-negativity after each update; symmetry was enforced at every iteration.
We applied gradient clipping with threshold $c=5$ and disabled weight decay ($\lambda=0$).
The learning rate started at $\alpha_0=0.04$ and was halved at epochs 700 and 2000 and additionally on plateau (patience of 7 logging blocks of 100 epochs), with $\alpha_{\min}=10^{-4}$.
Randomness was fixed by setting the optimizer seed to 42 for reproducibility.

We compare against:
\begin{itemize}[label=--]
\item \textbf{MW$^\star$-proj completion:} we infer an MW$^\star$ tree \cite{MakarenkovLapointe2004} from the incomplete distance matrix, and impute missing entries by the corresponding patristic distances.
\item \textbf{NJ$^\star$-proj completion:} we infer an NJ$^\star$ tree  \cite{SaitouNei1987} from the incomplete distance matrix (when the NJ$^\star$ steps are well-defined) and impute missing entries by the corresponding patristic distances.
\item \textbf{LRMC (Soft-Impute):} iterative Soft-Impute with singular-value soft-thresholding (fixed $\tau=0.02$, 100 iterations, tolerance $10^{-5}$), using a rank-$k$ truncated SVD ($k=10$) and enforcing symmetry and non-negativity by projection \cite{MazumderHastieTibshirani2010}.
\end{itemize}
All methods are constrained to keep the observed entries $D[\Omega_{p,r}]$ unchanged; only missing pairs are estimated.

$ $
\subsection{Evaluation protocol}
We define vectors
$\mathbf{x}=\{\widehat D^{(p,r)}_{ij}\}_{(i,j)\in\Omega_{\triangle}},$ and $
\mathbf{y}=\{D_{\mathrm{ref},ij}\}_{(i,j)\in\Omega_{\triangle}}$.

We report
\begin{align}
\mathrm{RMSE}&=\sqrt{\frac{1}{\binom{n}{2}}\sum_{(i,j)\in\Omega_{\triangle}}(\widehat D^{(p,r)}_{ij}-D_{\text{ref},ij})^2},\\
\mathrm{MAE}&=\frac{1}{\binom{n}{2}}\sum_{(i,j)\in\Omega_{\triangle}}\big|\widehat D^{(p,r)}_{ij}-D_{\text{ref},ij}\big|,
\end{align}
together with the Pearson correlation $r(\mathbf{x},\mathbf{y})$ and the Spearman rank correlation
$\rho = r\!\big(\operatorname{rank}(\mathbf{x}),\,\operatorname{rank}(\mathbf{y})\big).$

Also we compute and report $\overline{\Delta}(\widehat D^{(p,r)})$ from \eqref{eq:delta_normalized}.

To assess the downstream impact on phylogenetic structure, we additionally construct Neighbor-Joining (NJ) trees from $D_{\mathrm{ref}}$ and each completed matrix $\widehat D^{(p,r)}$ (after sanitization to enforce symmetry, non-negativity, and a zero diagonal). We then compare each inferred tree to the reference NJ tree using (i) the normalized Robinson--Foulds distance (topology), (ii) RMSE/MAE between patristic (leaf-to-leaf) distance matrices, and (iii) Pearson/Spearman correlations of patristic distances (all computed over $\Omega_{\triangle}$).

For readability, the main text reports only summary statistics and a small set of representative figures; replicate-level outputs (e.g.\ all NJ trees across methods and missingness levels) are provided in the Supplementary Material.

\section{Results and discussion}
Table~\ref{tab:main_results} compares Hyb-Adam-UM, MW$^\star$-proj, NJ$^\star$-proj, and LRMC across four missingness levels (30\%, 50\%, 65\%, and 85\%), reporting both matrix-level reconstruction metrics and downstream Neighbor-Joining (NJ) tree fidelity. Two consistent patterns emerge.

For moderate missingness (30--65\%), MW$^\star$-proj attains the strongest entrywise reconstruction accuracy (lowest RMSE/MAE and highest Pearson/Spearman correlations), while Hyb-Adam-UM achieves the smallest ultrametric-violation score $\overline{\Delta}(\widehat D)$ at every missingness level. This separation is expected: MW$^\star$ is optimized for distance-based phylogeny restoration, whereas Hyb-Adam-UM explicitly targets the triplet-based ultrametric objective. At extreme missingness (85\%), Hyb-Adam-UM becomes the most accurate method overall (best RMSE/MAE and correlations) while preserving a very small $\overline{\Delta}$, whereas MW$^\star$-proj and LRMC exhibit numerical instability in $\overline{\Delta}$ (large triangle-inequality violations under the robust penalty), indicating degraded global consistency.

In terms of topology, MW$^\star$-proj yields the lowest normalized Robinson--Foulds distance for 50--85\% missingness, while Hyb-Adam-UM is best at 30\% and remains competitive thereafter. For branch-length agreement (patristic distances), MW$^\star$-proj performs best for 30--65\% missingness (lowest patristic RMSE and highest patristic rank correlation), whereas Hyb-Adam-UM is best at 85\% and maintains the highest patristic Spearman at 65\% and 85\%. NJ$^\star$-proj is consistently fast but less accurate, and it fails at 85\% missingness (insufficient information in the partially observed matrix for a stable reconstruction in the tested setting).

In terms of runtime, the values in Table~\ref{tab:main_results} are reported for a reference single-core implementation to ensure consistent and reproducible timing across methods. NJ$^\star$-proj and LRMC are orders of magnitude faster (sub-second) but sacrifice accuracy and/or tree fidelity, while MW$^\star$-proj and Hyb-Adam-UM require tens of seconds per instance; Hyb-Adam-UM is typically faster than MW$^\star$-proj in these experiments but slower than the purely statistical baselines.  Hyb-Adam-UM repeatedly evaluates the triplet objective $\Delta(D)$ during optimization and, in particular, during finite-difference gradient estimation; a full evaluation of $\Delta(D)$ scales as $O(n^3)$ in the matrix size, and with $m$ missing entries the finite-difference loop yields a per epoch cost of $O(mn^3)$.  At the same time, the computational workload of Hyb-Adam-UM is highly parallelizable: finite-difference probes for different missing entries, along with the corresponding triplet evaluations, can be executed independently. This offers a clear route to substantial acceleration at larger scales; in our GPU implementation, imputing a \(100\times 100\) matrix on an RTX~5080 took \(15\text{--}20\) seconds.  Since the present study focuses on reconstruction quality and downstream tree fidelity, we report runtimes only as order-of-magnitude indicators rather than as optimized performance results.

\begin{table*}[t]
\caption{Combined matrix- and tree-level performance on $15\times 15$ mtDNA distance matrices (mean $\pm$ SD over 5 replicates per missingness level). RMSE/MAE and patristic RMSE are reported as $\times 10^{-2}$. Lower is better for RMSE/MAE, $\overline{\Delta}(\widehat D^{(p,r)})$, runtime, RF$_{\mathrm{norm}}$, and patristic RMSE; higher is better for correlations. Best values within each missingness level are in bold.}
\label{tab:main_results}
\centering
\scriptsize
\setlength{\tabcolsep}{3pt}
\renewcommand{\arraystretch}{1.12}
\begin{tabular}{llccccccccc}
\toprule
& & \multicolumn{6}{c}{Matrix metrics} & \multicolumn{3}{c}{Tree metrics} \\
\cmidrule(lr){3-8}\cmidrule(lr){9-11}
$p$ & Method & RMSE ($\times10^{-2}$) & MAE ($\times10^{-2}$) & Pear. & Spear. & $\overline{\Delta}(\widehat D^{(p,r)})$ & Time(s) & RF$_{\mathrm{norm}}$ & pat.\ RMSE ($\times10^{-2}$) & pat.\ Spear. \\
\midrule
\multirow{4}{*}{30\%} & Hyb-Adam-UM  & 0.75 $\pm$ 0.46 & 0.21 $\pm$ 0.10 & 0.97 $\pm$ 0.03 & 0.96 $\pm$ 0.04 & \best{0.18 $\pm$ 0.02} & 12.29 $\pm$ 0.07 & \best{0.37 $\pm$ 0.23} & 0.75 $\pm$ 0.47 & 0.95 $\pm$ 0.04 \\
& MW$^\star$-proj & \best{0.59 $\pm$ 0.53} & \best{0.15 $\pm$ 0.10} & \best{0.98 $\pm$ 0.03} & \best{0.97 $\pm$ 0.03} & 0.21 $\pm$ 0.02 & 29.40 $\pm$ 2.60 & 0.43 $\pm$ 0.25 & \best{0.59 $\pm$ 0.53} & \best{0.97 $\pm$ 0.04} \\
& NJ$^\star$-proj   & 1.03 $\pm$ 0.68 & 0.30 $\pm$ 0.25 & 0.95 $\pm$ 0.05 & 0.91 $\pm$ 0.08 & 0.30 $\pm$ 0.18 & \best{ $<$0.01} & 0.45 $\pm$ 0.10 & 0.99 $\pm$ 0.64 & 0.91 $\pm$ 0.07 \\
& LRMC  & 2.98 $\pm$ 0.17 & 1.41 $\pm$ 0.09 & 0.76 $\pm$ 0.03 & 0.74 $\pm$ 0.03 & 1.79 $\pm$ 0.25 & 0.07 $\pm$ 0.00 & 0.78 $\pm$ 0.10 & 2.77 $\pm$ 0.31 & 0.70 $\pm$ 0.09 \\
\midrule
\multirow{4}{*}{50\%} & Hyb-Adam-UM  & 1.10 $\pm$ 0.43 & 0.38 $\pm$ 0.12 & 0.95 $\pm$ 0.03 & 0.92 $\pm$ 0.04 & \best{0.12 $\pm$ 0.02} & 19.74 $\pm$ 0.09 & 0.47 $\pm$ 0.07 & 1.07 $\pm$ 0.46 & 0.92 $\pm$ 0.04 \\
& MW$^\star$-proj  & \best{0.93 $\pm$ 0.50} & \best{0.29 $\pm$ 0.11} & \best{0.96 $\pm$ 0.03} & \best{0.94 $\pm$ 0.04} & 0.17 $\pm$ 0.02 & 28.23 $\pm$ 5.03 & \best{0.38 $\pm$ 0.13} & \best{0.94 $\pm$ 0.49} & \best{0.94 $\pm$ 0.04} \\
& NJ$^\star$-proj   & 2.25 $\pm$ 0.77 & 0.93 $\pm$ 0.62 & 0.83 $\pm$ 0.08 & 0.80 $\pm$ 0.12 & 4.79 $\pm$ 8.39 & \best{ $<$0.01} & 0.57 $\pm$ 0.14 & 2.19 $\pm$ 0.78 & 0.79 $\pm$ 0.11 \\
& LRMC  & 4.59 $\pm$ 0.22 & 2.76 $\pm$ 0.11 & 0.55 $\pm$ 0.04 & 0.51 $\pm$ 0.06 & 2.88 $\pm$ 0.63 & 0.07 $\pm$ 0.00 & 0.80 $\pm$ 0.14 & 4.34 $\pm$ 0.11 & 0.43 $\pm$ 0.16 \\
\midrule
\multirow{4}{*}{65\%} & Hyb-Adam-UM  & 1.91 $\pm$ 1.10 & 0.64 $\pm$ 0.22 & 0.87 $\pm$ 0.11 & \best{0.90 $\pm$ 0.05} & \best{0.10 $\pm$ 0.06} & 25.80 $\pm$ 0.12 & 0.53 $\pm$ 0.05 & 1.41 $\pm$ 0.54 & \best{0.90 $\pm$ 0.04} \\
& MW$^\star$-proj  & \best{1.40 $\pm$ 0.84} & \best{0.63 $\pm$ 0.38} & \best{0.90 $\pm$ 0.11} & 0.89 $\pm$ 0.08 & 0.18 $\pm$ 0.03 & 30.53 $\pm$ 3.82 & \best{0.42 $\pm$ 0.12} & \best{1.39 $\pm$ 0.83} & 0.88 $\pm$ 0.08 \\
& NJ$^\star$-proj  & 2.48 $\pm$ 0.42 & 1.23 $\pm$ 0.40 & 0.78 $\pm$ 0.08 & 0.76 $\pm$ 0.07 & 0.66 $\pm$ 0.10 & \best{ $<$0.01} & 0.70 $\pm$ 0.05 & 2.38 $\pm$ 0.40 & 0.77 $\pm$ 0.07 \\
& LRMC  & 6.12 $\pm$ 0.54 & 4.32 $\pm$ 0.41 & 0.37 $\pm$ 0.08 & 0.34 $\pm$ 0.11 & 3.26 $\pm$ 0.44 & 0.08 $\pm$ 0.00 & 0.92 $\pm$ 0.10 & 5.56 $\pm$ 0.64 & 0.30 $\pm$ 0.16 \\
\midrule
\multirow{4}{*}{85\%} & Hyb-Adam-UM  & \best{3.85 $\pm$ 0.43} & \best{2.21 $\pm$ 0.32} & \best{0.39 $\pm$ 0.14} & \best{0.37 $\pm$ 0.17} & \best{0.02 $\pm$ 0.01} & 33.78 $\pm$ 0.13 & 0.88 $\pm$ 0.10 & \best{3.84 $\pm$ 0.43} & \best{0.37 $\pm$ 0.17} \\
& MW$^\star$-proj  & 4.19 $\pm$ 0.37 & 2.71 $\pm$ 0.31 & 0.32 $\pm$ 0.06 & 0.27 $\pm$ 0.11 & $\approx 10^{10}$ & 29.38 $\pm$ 3.60 & \best{0.87 $\pm$ 0.10} & 4.20 $\pm$ 0.37 & 0.27 $\pm$ 0.12 \\
& NJ$^\star$-proj   & N/A & N/A & N/A & N/A & N/A & N/A & N/A & N/A & N/A \\
& LRMC  & 10.23 $\pm$ 0.60 & 8.64 $\pm$ 0.58 & 0.10 $\pm$ 0.11 & 0.04 $\pm$ 0.15 & $\approx 10^{10}$ & \best{0.07 $\pm$ 0.00} & 0.98 $\pm$ 0.04 & 8.78 $\pm$ 0.41 & 0.00 $\pm$ 0.11 \\
\bottomrule
\end{tabular}
\vspace{2pt}
\parbox{\textwidth}{\scriptsize Runtime is reported as mean $\pm$ SD over the 5 replicates at each missingness level (NJ$\star$ uses successful replicates only). $\overline{\Delta }(D_{\mathrm{ref}})= 0.23 $.}
\end{table*}

\begin{figure*}[t]
  \centering
  \includegraphics[width=\textwidth]{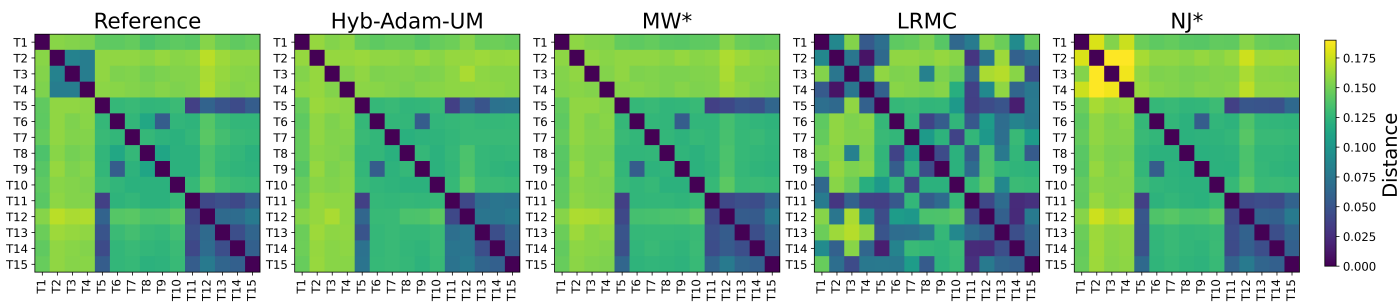}
  \caption{Heatmaps of the reference distance matrix $D_{\mathrm{ref}}$ and the reconstructed matrices for a representative replicate at $p=50\%$ missingness.}
  \label{fig:heatmap_p50}
\end{figure*}

\begin{figure*}[t]
  \centering
  \includegraphics[width=\textwidth]{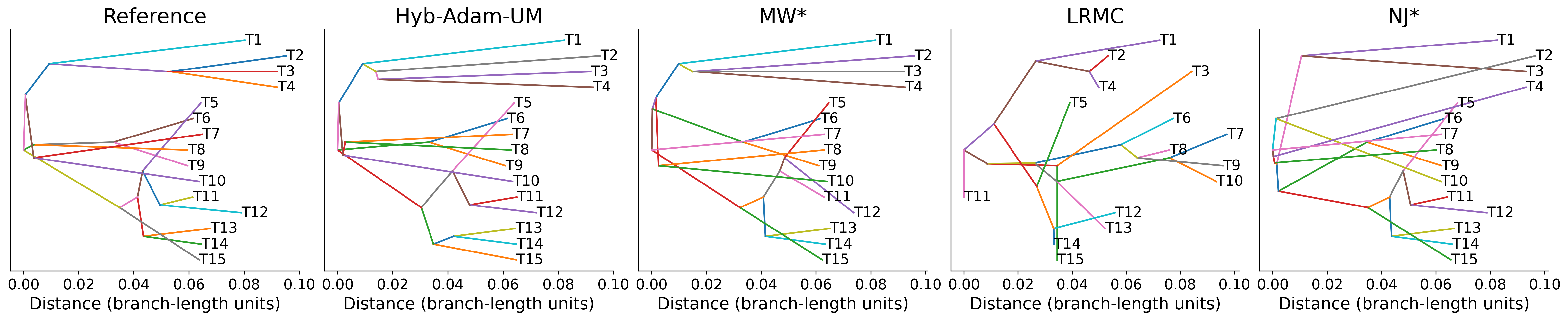}
  \caption{Neighbor-Joining trees inferred from the reference matrix $D_{\mathrm{ref}}$ and from the reconstructed matrices shown in Figure~\ref{fig:heatmap_p50} (same representative replicate at $p=50\%$ missingness).}
  \label{fig:njtrees_p50}
\end{figure*}

For qualitative inspection, we visualize heatmaps of the reference matrix and the reconstructed matrices for a representative replicate at missingness level $p=50\%$ (Figure~\ref{fig:heatmap_p50}).

We additionally show the corresponding Neighbor-Joining trees inferred from the same reference and reconstructed matrices in the subsequent figure (Figure~\ref{fig:njtrees_p50}; see below). 
Complete replicate-level heatmaps and trees for all methods and missingness levels are archived in a public repository (see Data availability).

\subsection{Optimization dynamics}

\begin{figure}
  \centering
  \includegraphics[width=\columnwidth]{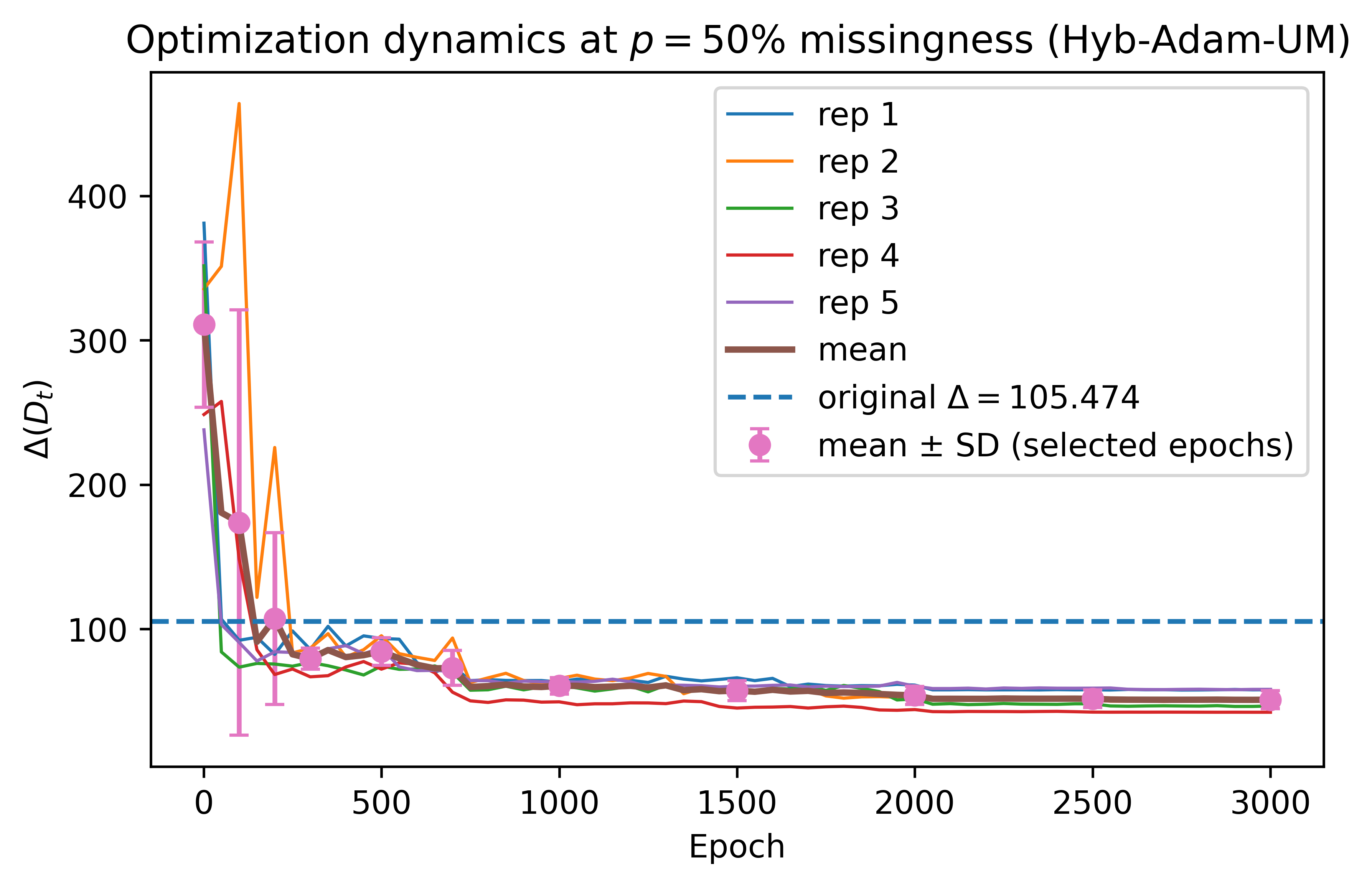}
  \caption{Optimization dynamics of Hyb-Adam-UM at $p=50\%$ missingness: $\Delta(D_t)$ versus epoch for five masking replicates (lines). Circles with error bars indicate mean $\pm$ SD across replicates at selected epochs. The dashed horizontal line shows the baseline $\Delta(D_{\mathrm{ref}})=105.474$ for the reference complete matrix.}
  \label{fig:delta_dynamics_p50}
\end{figure}

To illustrate the optimizer behavior, we track the global violation score $\Delta(D_t)$ over epochs.
For the $p=50\%$ missingness setting, Figure~\ref{fig:delta_dynamics_p50} shows the trajectories across the five masking replicates, together with error bars indicating mean $\pm$ SD of $\Delta(D_t)$ at selected epochs.
For reference, we also plot the baseline value $\Delta(D_{\mathrm{ref}})=105.474$ of the original (complete) matrix derived from Needleman--Wunsch alignments.
Notably, in our runs $\Delta(D_t)$ drops below $\Delta(D_{\mathrm{ref}})$, indicating improved ultrametric consistency relative to the raw Needleman--Wunsch--derived distances in terms of the proposed score.

Overall, $\Delta(D_t)$ decreases substantially during optimization, supporting the practical feasibility of Adam-style updates for a non-smooth, triplet-based objective.

We quantify the improvement relative to the initialized matrix $D_{\mathrm{init}}$ (the starting point of the optimization) by
\[
\mathrm{Reduction}(\%)=\frac{\Delta(D_{\mathrm{init}})-\Delta(\widehat D)}{\Delta(D_{\mathrm{init}})}\times 100\%.
\]
For $p=50\%$ missingness, we compute this reduction for each of the five masking replicates and then summarize across replicates. On average, $\Delta$ decreases from $346.13\pm 62.66$ at initialization to $53.67\pm 20.56$ after optimization, corresponding to an average reduction of $84.05\%\pm 5.97\%$ (mean $\pm$ SD over 5 replicates).

\section{Conclusion}

We proposed Hyb-Adam-UM, a hybrid approach to completing mtDNA distance matrices that combines (i) a biologically informed warm start from a limited set of Needleman--Wunsch alignments with (ii) ultrametric-aware refinement via Adam-style minimization of a robust triplet-violation objective $\Delta(D)$ while keeping observed distances fixed. In controlled experiments on $15\times 15$ mtDNA matrices with 30\%, 50\%, 65\% and 85\% missingness (five masking replicates per level), Hyb-Adam-UM consistently enforces strong ultrametric consistency across all missingness levels. In terms of entrywise reconstruction and downstream Neighbor-Joining fidelity, the method is competitive with MW$^\star$-proj, NJ$^\star$-proj, and LRMC (Soft-Impute) at moderate missingness (30--65\%), and it becomes the most accurate approach at extreme missingness (85\%), where projection- and low-rank baselines exhibit degraded global consistency or fail to produce stable reconstructions. Overall, these results indicate that explicitly optimizing ultrametric structure is most beneficial in the highly sparse regime, improving both matrix restoration and the reliability of inferred phylogenetic trees when only a small fraction of pairwise distances can be computed.

\section*{Funding}
This work was supported by the National Key Research and Development Program of China (Grant No. 2025YFE0113400), Shenzhen Science and Technology Program (No. RCJC\linebreak 20231211090030059), National Key Research and Development Program of China (No. W2421102) and the Scientific Program of Chinese Universities ``Higher Education Stability Support Program'' (No.\ 20220819092934001).

\section*{Competing interests}
The first author is an inventor on a granted Chinese invention patent related to the method described in this manuscript (Patent No.\ ZL 202510230572.7; grant date: 17~Oct~2025). The patent is assigned to Shenzhen MSU--BIT University. The authors declare no other competing interests.

\section*{Data availability}
To support reproducibility, the reference distance matrix, all masking patterns, completed matrices, and the full evaluation scripts (including tree reconstruction and metrics) are available on GitHub: \url{https://github.com/mitichya/hyb-adam-um/}. The exact version used in this paper is archived on Zenodo: \url{https://doi.org/10.5281/zenodo.18609747}.

% ==================== References ====================
% Self-contained references via thebibliography. Replace/extend with your .bib if preferred.
%UNCOMMENT THE BELOW TWO LINES IN CASE YOU NEED NUMBERED FORMAT.
%\bibliographystyle{oup-plain}
%\bibliography{reference}

%UNCOMMENT THE BELOW TWO LINES IN CASE YOU NEED AUTHOR YEAR FORMAT.
%\bibliographystyle{oup-abbrvnat}
\bibliographystyle{unsrtnat}
\bibliography{reference}

%\bibliographystyle{unsrt}
%\bibliography{reference.bib}

\end{document}